# Towards a Joint Understanding of Remote Operation for Vehicles in Public Road Traffic

Elisabeth Shi [1*], Maria-Magdalena Wolf [2], Nina Theobald [3], Bettina Abendroth [3], Eugen Wige [4], Johannes Springer [5], Katharina Hottelart [4], Andreas Schrank [6], Thorben Brandt [6], Michael Oehl [6], Frank Diermeyer [2], & Lena Plum [1]

[1] Section F4 Automated Driving, Federal Highway and Transport Research Institute (BASt), Bruederstr. 53, D - 51427 Bergisch Gladbach, Germany; plum@bast.de

[2] Institute of Automotive Technology, Technical University of Munich (TUM), Boltzmannstr. 15, DE-85748 Garching bei München, Germany; maria.wolf@tum.de, diermeyer@tum.de

[3] Technical University of Darmstadt, Department of Mechanical Engineering, Institute of Ergonomics and Human Factors, Otto-Berndt-Straße 2, 64287 Darmstadt, Germany; nina.theobald@tu-darmstadt.de, bettina.abendroth@tu-darmstadt.de

[4] Valeo Schalter und Sensoren, Driving Assistance Product Group, Hummendorfer Str. 74, 96317 Kronach Neuses, Germany; eugen.wige@valeo.com

[5] Digital Division – Connected Mobility, T-Systems International GmbH, Deutsche Telekom AG, Holzhauser Str. 4-8, 13509 Berlin, Germany; johannes.springer@posteo.de

[6] Institute of Transportation Systems, German Aerospace Center (DLR), Lilienthalplatz 7, 38108 Braunschweig, Germany; andreas.schrank@dlr.de, thorben.brandt@dlr.de, michael.oehl@dlr.de,

* Correspondence: elisabeth.shi@th-rosenheim.de

## Abstract

Sustained driving automation systems are envisioned to be used as the foundation for driverless mobility services. However, both researchers and practitioners acknowledge that current driving automation systems are not yet able to handle all traffic situations that a human driver can handle. To bridge this gap and enable mobility services without an in-vehicle human driver or fallback, remote operation (or teleoperation) is increasingly discussed. Recently, first legal actions have been taken to enable some forms of remote operation on public roads. Remote operation encompasses a broad spectrum of methods to support a driving automation system, ranging from remote assistance, which includes providing information or releasing a maneuver, to remote driving, which includes driving the vehicle from a remote location. As such, safe implementation of remote operation in public road traffic challenges the collaboration of multiple academic disciplines (e.g. engineering, psychology, informatics, law, etc.) and stakeholders (e.g. remote operation service providers, remote operators, vehicle manufacturers, regulatory authorities, etc.). At the same time, the interdisciplinary discourse is often challenging due to differing expectations and language. To build a common ground, this article traces terminology back to the original differences in information processing both on human and vehicle side. This framework aims to help further discourse by directly specifying what is needed to engage a diverse audience including researchers and stakeholders of different backgrounds and interests. Recently discussed forms of teleoperation are integrated into this framework.

# 1 Introduction

Researchers and practitioners agree that it is unrealistic for current driving automation systems (DAS) to handle all traffic situations that in-vehicle human drivers can manage today (Amador et al., 2022; Bogdoll et al., 2022; Brecht et al., 2024; Prakash, Vignati, & Sabbioni, 2023). Rather, DAS are designed to perform the dynamic driving task (DDT, definition 3.10 in SAE J3016, 2021) under specific conditions. The DDT includes the lateral and longitudinal vehicle motion control, the object and event detection (including recognition and classification) and response (including preparation and execution) (OEDR), as well as on a tactical level maneuver planning and enhancing conspicuity as needed. SAE J3016 (2021) classifies DAS according to the extent to which they can perform the DDT on a sustained basis (Level 0 to Level 5, definitions 5.1 to 5.6 in SAE J3016, 2021). The focus on sustained driving automation excludes other systems such as emergency braking systems or forward collision warnings (section 1 in SAE J3016, 2021). The conditions under which a DAS is designed to perform the DDT are referred to as the operational design domain (ODD, definition 3.21 in SAE J3016, 2021) which defines a DAS's operational limits. When DASs are used to provide a driverless transportation service (for passengers or freight), it is necessary to compensate for ODD limits that occur during a trip. In this context, a remotely located person can perform different tasks to support a DAS at its ODD limits via a wireless communication network. There is ongoing discourse regarding the nature, scope and definition of these tasks (Amador et al., 2022; Prakash, Vignati, & Sabbioni, 2023; Wolf et al., 2025). Remote operation is seen as a bridging technology for DASs and as an enabler for driverless transportation services in the near future (Bogdoll et al., 2022; Brecht et al., 2024; Zhao et al., 2023).

In recent years, research activities have intensified on the topic of remote operation of on-road vehicles (Aramrattana et al., 2025; Brandt et al., 2024; Majstorovic et al., 2022; Schrank et al., 2024; Schrank et al., 2025; Schwindt et al., 2023; Schwindt-Drews & Abendroth, 2025; Schwindt-Drews et al., 2025; Walocha et al., 2025). Their practical implementation and the development of a legal framework run in parallel. However, inconsistent terminology creates communication barriers and fragments knowledge within scientific discourse and practical application. Moreover, it complicates education and training for remote operation use cases and impedes efforts toward standardization and policy formulation. Addressing these challenges is key for enhancing clarity, enabling effective collaboration and advancing understanding and development within the field of remote operation.

## 1.1 Terminology and Scope of this Article

Regarding terminology on automated driving, this article is based on SAE International Standard J3016 (2021). SAE J3016 applies the term "driving automation system" (DAS, definition 3.6 in SAE J3016, 2021) to refer to systems from Level 1 to 5. In contrast, the term "automated driving systems" (ADS, definition 3.2 in SAE J3016, 2021) is exclusive to systems of Level 3 to 5. The term DAS is used throughout the article to be more generic because SAE Levels 1 and 2 are not excluded from our proposed framework. Unless otherwise stated, in this article, the terms DAS and ADS always refer to the in-vehicle available systems.

Both “remote operation” (e.g. Amador et al., 2022) and “teleoperation” (e.g. Prakash, Vignati, & Sabbioni, 2023) are commonly used to describe the same concept (e.g. Amador et al., 2022; Tener & Lanir, 2023; Zhao, 2023). The term “teleoperation” combines the Greek prefix “tele” (remote) with the Latin “operatio” (operation), making the interchangeable use of both terms understandable. This work treats both terms as synonyms and predominantly uses “remote operation”. This also applies to semantically related terms, such as “remote operator” or “teleoperator”. To avoid confusion, it should be noted that in robotics, “teleoperator” refers to the remotely operated robot (here, the vehicle). In on-road vehicle research, however, the term is used to denote the person or entity performing the remote operation from a distance. Although remote operation does not inherently require a human operator (e.g. Amador et al., 2022; Bogdoll et al., 2022; Majstorovic et al., 2022), it is currently understood to refer to human-operated control from a distance. While future concepts may involve an automated remote operator (Chehwan, 2025), this article follows the prevailing interpretation of a human operator.

Research and practice consider that a remotely located person may perform tasks that are related to vehicle motion control (e.g. actively performing the DDT and monitoring the vehicle and its surroundings, or providing requested information regarding an unidentified object detected by the DAS) as well as tasks that are not directly related to vehicle motion control (e.g. communication with in-vehicle passengers, or coordination tasks like fleet management, Bogdoll et al., 2022; Schwindt et al., 2023; Wolf et al., 2025). Within the multitude of conceivable tasks that a person can perform from a remote location, this article focuses on activities that target vehicle motion control in public road traffic, and the spectrum of interactions between the remote operator and the DAS available in the vehicle. This article follows the notion that remote operation of any type is an auxiliary means to provide driverless rides.

## 1.2 Aim

The authors propose a joint understanding and terminology for concepts of remote operation that is applicable and useable for purposes in the highly interdisciplinary field, including diverse academic disciplines (e.g. engineering, psychology, informatics, law,…) and stakeholders (e.g. remote operation service providers, remote operators, vehicle manufacturers, regulatory authorities,…). Our work is based on (1) a review of currently used terminology and existing standards, (2) the development of a framework outlining both human and technical information processing in remote operation, and (3) the derivation of definitions for remote assistance and remote driving and their demarcation.

Remote assistance and remote driving, as two major types of remote operation, are distinguished based on their underlying information flow between human and vehicle information processing. This allows to highlight the key differences of these two approaches to remote operation and their respective interaction with sustained DAS of different SAE levels.

# 2 Literature Review on Terminology and Standards

This section first gives an overview of concepts and applications of remote operation considered in recent activities in research, standardization and legal context, including the terminology used to refer to them.

In a review of pertinent literature, Schwindt et al. (2023) reveal varying approaches to categorizing and labeling teleoperation tasks. These differ primarily in the number of task categories utilized and the terminology employed for these categories. Additionally, the review shows that on the one hand, different tasks are grouped under the same terms, and on the other hand, different terms are used for the same tasks. There are also variations in how the remote operators' tasks are interpreted in detail. Mostly the term “remote operation” is used as the top-level term that summarizes a set of more specific tasks. BSI Flex 1887 (2024) uses the term “remote operation” as the top-level term while the technical report by 5GAA (2021) applies the term “teleoperation”. SAE J3016 (2021) does not introduce an overarching category for remote activities it defines, and explicitly refrains from using the term “teleoperation” arguing that it “is not defined consistently in the literature” (SAE J3016, 2021, p. 19). However, more recent publications (after SAE J3016's last update) still address and apply the term “teleoperation” (e.g. Majstorovic et al., 2022; Prakash, Vignati, Vignarca, et al., 2023; Schwindt et al., 2023; Working Group “Research Needs in Teleoperation”, 2025; Zhao, 2023). The informal document discussed in UNECE WP 1 (2024) uses the term “remote management” to describe remote driving, remote monitoring and assistance. In contrast, Bogdoll et al. (2022) apply the same term “remote management” as a top-level term for all activities not related to driving, such as “dispatching or interaction with passengers” (Bogdoll et al., 2022, p. 11). This illustrates that the same term may refer either to DDT-related remote activities or to activities outside the DDT-related scope of this article. Since this article focusses on activities that target vehicle motion control and does not include activities not related to driving, this article follows the predominant notion and uses “remote operation” as the more appropriate top-level term summarizing these specific tasks.

Remote operation is frequently categorized into three types, typically distinguishing between remote driving, remote assistance and remote monitoring (Amador et al., 2022; Bogdoll et al., 2022; Majstorovic et al., 2022). According to Bogdoll et al. (2022), “remote driving” describes the case when the remote operator “directly controls the vehicle in the form of steering and throttle/brake commands.” (Bogdoll et al., 2022, p. 10). Bogdoll et al. (2022) regard teleoperation in its strict sense as equivalent to remote driving. Furthermore, they describe “Remote assistance“ as the case when the remote operator provides information or authorizations to the vehicle's ADS (Bogdoll et al., 2022). This may include supporting the ADS' OEDR, e.g. classification of objects, interpretation of other road users' hand gestures, or authorizing one or several maneuvers suggested by the ADS. Note that ADS are capable of performing the DDT on their own within their ODD, excluding lower levels of driving automation that require a human driver to perform at least part of the DDT. “Remote monitoring” does not include direct interaction with vehicle motion control, and is therefore not considered a separate category in other frameworks (5GAA, 2021; Working Group “Research Needs in Teleoperation”, 2025). Instead, it is understood as the prerequisite for remote driving and remote assistance and refers to the observation and assessment of data received from the ADS (Bogdoll et al., 2022). Such data can be “anything” (Bogdoll et al., 2022, p. 10), e.g. information on the vehicle's environment, location, or state. For the scope of the present article, remote monitoring is therefore relevant as an enabling activity that may precede or accompany remote assistance or remote driving, but is not treated as a separate form of remote operation.

Wolf et al. (2025) further distinguish “remote intervention” as a use case where the remote operator directly influences a vehicle's “movement, status or conspicuity” (p. 2540), overriding the ADS without remote driving or assistance. Remote intervention includes releasing an ADS-equipped vehicle to start or continue its ride automatically, initiating a safe stop, or deactivating the ADS or vehicle. Conspicuity can be enhanced by using the vehicle's horn or lights. Wolf et al. (2025) highlight that remote intervention differs from remote driving in that it does not include activities related to the sustained performance of the DDT, e.g. steering or using turn signals.

SAE J3016 (2021) provides definitions for "remote assistance" (definition 3.23 in SAE J3016, 2021) and "remote driving" (definition 3.24 in SAE J3016, 2021): Remote assistance constitutes "*Event-driven provision, by a remotely located human (…), of information or advice to an ADS-equipped vehicle in driverless operation in order to facilitate trip continuation when the ADS encounters a situation it cannot manage.*" (SAE J3016, 2021, p. 18) Remote driving constitutes "*Real-time performance of part or all of the DDT* [dynamic driving task] *and/or DDT fallback (including, real-time braking, steering, acceleration, and transmission shifting), by a remote driver*." (SAE J3016, 2021, p. 19). While SAE J3016 (2021) does not provide specification of the remote assistant's location relative to the vehicle, it explicitly allows for a remote driver to be physically located inside the vehicle, provided that the person is not seated in the driver's seat and does not use the built-in control elements (definition 3.31.1.2 in SAE J3016, 2021). This differs from other approaches, such as the Working Group "Research Needs in Teleoperation" (2025), BSI Flex 1887 (2024), 5GAA (2021) and the informal document discussed in UNECE WP 1 (2024), which define the remote driver as being located outside the vehicle.

The Working Group "Research Needs in Teleoperation" (2025) extends SAE J3016's definitions on remote driving and assistance to include the criterion that the remotely located person is without direct view of the vehicle. In addition, remote driving is further subdivided into two application use cases: Continuous remote driving and event-based remote driving. "Continuous remote driving" describes the use case in which an entire trip is executed through remote driving only. No ADS is active or available in the vehicle and no in-vehicle driver is present or available. Continuous remote driving is considered outside the scope of this work (section 1.1), as the focus is placed on remote operation as an auxiliary means of enabling driverless rides. "Event-based remote driving" describes the use case in which a vehicle is operated by an ADS of SAE Level 4 or 5, and encounters a situation that the ADS cannot solve. The ADS requests remote driving which is then executed by the remote driver for a limited period of time, and/or for a specific section of the trip only. After that, the remote driver hands back the driving task to the Level 4 or 5 ADS, which continues the trip.

5GAA (2021) also builds on SAE J3016's definitions but uses different terms: "Indirect Control ToD [Tele-operated Driving]" corresponds to SAE J3016's remote assistance, and "Direct Control ToD" corresponds to remote driving. 5GAA's (2021) definitions are expanded to also include strategic operations, such as selecting a new route, if needed. It should be noted that, in line with the scope defined in section 1.1, the article at hand focusses on tasks immediately related to the execution of the DDT, thereby leaving strategic tasks outside this article's scope.

The reviewed literature shows that existing approaches to remote operation differ not only in terminology, but also in the categorization of remote operation and the criteria used to distinguish between specific forms of remote operations. Nevertheless, several recurring dimensions can be identified across research, standardization and regulatory definitions. These include, for example, the trigger, initiator and frequency of remote operator involvement, the extent to which the remote operator contributes to or performs the DDT, the type of support provided to the vehicle or DAS (e. g. information, advice, authorization or direct control), the role of remote monitoring and the remote operator's location relative to the vehicle.

Rather than introducing additional dimensions as the primary basis for distinguishing between remote assistance and remote driving, this article derives the distinction from the interaction between the remote operation control center and the vehicle's DAS, with particular emphasis on how information provided by the remote operation control center is processed within the vehicle-side DAS architecture. This perspective is motivated by the observation that the dimensions identified in the literature are frequently considered in isolation and are not consistently linked to the interaction between the remote operation control center and the vehicle's ADS, despite the fact that remote operation fundamentally represents such an interaction. The dimensions identified above are subsequently employed to characterize and compare the resulting definitions of remote assistance and remote driving.

# 3 A Joint Understanding of Remote Operation

To illustrate the process of remote operation and clearly distinguish the concepts of remote assistance and remote driving, a framework was developed. The developed framework (Figure 1) consists of the vehicle with its DAS and remote operation modules to be operated on the left side, the remote operation control center with the remote operator and remote operator workplace on the right side, and overarching external information sources such as weather conditions or passenger input. A wireless communication network connects the vehicle and the remote operation control center. The framework illustrates the abstract information-processing activities of both the DAS and the remote operator, with information exchanged and, if necessary, modified through the remote operation modules and the remote operator workplace before being transmitted via the network. This process is supported by dedicated safety concepts on the side of the vehicle, the network, and the remote operation control center. The remote operator and the DAS pursue the shared objective of executing the DDT, with the remote operator complementing and supporting the DAS in the event of failures or uncertainties in the technical information-processing.

The following subsections first focus on each side, i.e. the remote operation control center and the vehicle, describing the respective information processing and components depicted in Figure 1. Next, the delineation between remote driving and remote assistance is described based on the interaction of information processing of both sides.

## 3.1 Remote Operation Control Center Components

The *remote operation control center* encompasses the *remote operator*, the *remote operator workplace*, and the *safety concept* that safeguards remote operation by influencing factors within the control center (see right side of Figure 1).

### 3.1.1 Remote Operator: Human Information Processing

The *remote operator* processes information and performs actions as e.g. described in Parasuraman et al.'s (2000) model of information processing. This model has been particularly designed for application in the context of human-machine interaction, making it highly suitable for this article's objective of human and vehicle interaction in remote operation. However, it is important to acknowledge that the model does not fully capture the complexity of human information processing.

In Parasuraman et al.'s (2000) model, *sensory processing* constitutes the first stage of information processing. This stage includes *sensory perception* and *selective attention* processes prior to *conscious perception*, such as orienting gaze movement for visual perception. In the second stage, information is consciously perceived. In Figure 1, sensation and perception are summarized in one processing stage labelled “sense”. Following Parasuraman et al.’s (2000) model, information is then manipulated in *working memory* to prepare *decision-making* performed in the third stage. Based on this decision, in the fourth stage *response selection and execution* takes place. In Figure 1, processes following sensation and perception and preceding response execution are summarized in the “plan” stage, while the “act” stage includes response execution. Such three-staged human information processing is also described by Bubb et al. (2015).

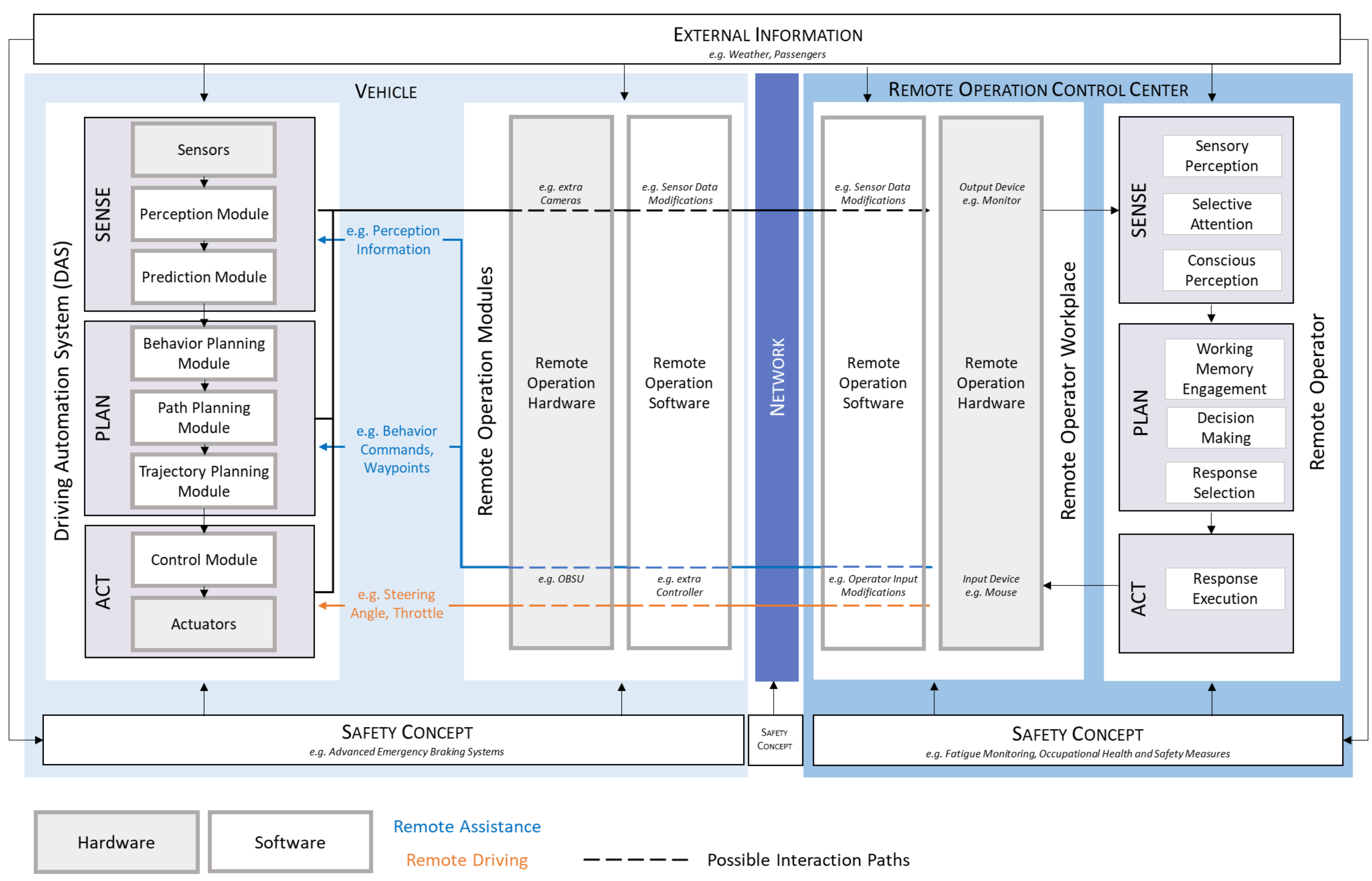

***Figure 1:** Delineation of remote assistance and remote driving based on integration of technical and human information processing in the remote operation system.*

### 3.1.2 Remote Operator Workplace

The remote operation control center contains a *workplace* to enable the remote operator to interact with the vehicle. The remote operator workplace consists of *hardware* and *software*. Incoming information from the vehicle may be modified by software (e.g. modifications of the sensor data) and displayed by hardware (output devices, e.g. monitors) to the human remote operator. Also, the remote operator workplace sends information to the vehicle. The remote operator uses hardware (input devices, e.g. mouse) to generate input for the vehicle’s DAS. This operator input may be modified by software, e.g. entered waypoints that are connected to form a path, before being sent to the vehicle side.

### 3.1.3 Safety Concept on the Remote Operation Control Center Side

Within the remote operation control center, a *safety concept* safeguards remote operators’ work and working conditions to allow for safe remote operation. This safety concept may include i.a. technical systems (e.g. systems that monitor remote operators’ fatigue), but also work and workplace related processes and measures that ensure remote operators’ qualification and occupational health and safety.

## 3.2 Vehicle Components

The vehicle encompasses the *driving automation system*, *remote operation modules* and a *safety concept* that safeguards the (remote) vehicle operation by influencing factors within the vehicle (see left side of Figure 1).

### 3.2.1 Driving Automation System: Technical Information Processing

Modeled after the human information processing model, the structure of the DAS can be grouped into three categories: *Sense*, *Plan*, and *Act* (Majstorovic et al., 2022). Apart from variations across different implementations, driving automation software consists of distinct modules within the vehicle, each performing specific tasks in a hierarchical structure. This publication uses Karle and Lienkamp’s (2021) broadly applicable and generalistic division of the automation software into six modules. Depending on the use case and implementation, not all (or even no) DAS modules are available or active.

The *Sense* block starts with receiving input from *sensors*, such as cameras or LiDARs. The sensor input data is then analyzed further in the *perception module*. First, the position and orientation of the ego-vehicle are computed through mapping and localization. At the same time, surrounding objects are clustered, identified, and additional features are extracted from the sensor data by a perception algorithm to derive an abstract environmental model. Afterwards, the *prediction module* estimates future states, respectively locations, of the perceived objects using the trajectories of other road participants. This data on the perception of the ego-vehicle, other surrounding objects, and their predicted states serves as input for the planning process.

The planning process (*Plan*) can be divided into two different levels: global path planning and local behavior/trajectory planning (Teng et al., 2023). Global path planning serves for navigating from the origin to the desired destination and forms the basis for local behavior/trajectory planning. Global path planning is not further considered in this publication, as the focus lies on the execution of the DDT. Local behavior/trajectory planning generates a valid and safe short-term trajectory based on the processed sensor information to reach the given global path. To explain the remote operator's interactions with the planning module in detail, this publication further separates local behavior/trajectory planning into the *behavior planning module*, *path planning module* and *trajectory planning module* (Karle & Lienkamp, 2021; Majstorovic et al., 2022). The *behavior planning module* chooses a predefined action, such as waiting or overtaking, to reach the given global path, while the *path planning module* generates a concrete collision-free path according to the selected action. The *trajectory planning module* creates the corresponding velocity profile to follow the entered local path.

In the *Act* block, the *control module* takes car of following the planned trajectory and regulates the velocity and position to guarantee a smooth and safe trip. Finally, the data is given to the *actuators* which include steering, throttle, and brake so that the vehicle can execute the lateral and longitudinal driving tasks. The DAS continuously monitors its environment and adjusts its behavior, ultimately controlling the vehicle's actuators.

### 3.2.2 Remote Operation Modules

The *remote operation modules* enable the vehicle's DAS to interact with the remote operation control center. The remote operation modules can consist of additional *hardware* and *software* components specific to the remote operation tasks. For example, remote operation might need extra cameras for additional views (hardware) for the remote operator or the stitching of existing camera views (software), which represent information provided by the vehicle to the remote operation control center. The other way round, an example for incoming information from the remote operation control center via the remote operation modules could be a dedicated remote operation controller on the vehicle side (software), which can prevent inappropriate behavior such as steering inputs that may lead to collisions, or additional computing power provided by an On-Board Safety Unit (OBSU, hardware). Also time synchronization mechanisms between remote operation control center and vehicle and respective failure mode executions might be part of the OBSU.

### 3.2.3 Safety Concept on the Vehicle Side

The *safety concept* on the vehicle side represents how functions contribute to safeguarding the remote operation use case (rather than safeguarding an in-vehicle human driver). This includes preventing the vehicle from entering a hazardous situation or exiting the DAS's ODD, ensuring a safe and comfortable stop in the event of a connection loss, limiting the maximum allowable speed based on the proximity of surrounding objects, or monitoring the vehicle diagnostic states and to pro-actively avoid the independent safety mechanisms such as Advanced Emergency Braking (AEB). The specific functions integrated in the *safety concept* depend on the use case in which remote operation is employed.

## 3.3 Interaction of Technical and Human Information Processing

The preceding subsections described the two sides of the remote operation system separately. The following subsection derives the distinction between remote assistance and remote driving from their interaction.

In remote operation, the DAS's information processing is either continuously supported or fully replaced by human information processing, or it receives support whenever modules in the technical information processing chain fail or produce uncertain results. By integrating the technical and human information processing pathways and examining the flow of information between the *vehicle* and the *remote operation control center*, the concepts of remote assistance and remote driving can be clearly distinguished.

### 3.3.1 Flow of Information and its Modification

Remote operation requires an exchange of information between the vehicle and the remote operation control center. The remote operation modules in the vehicle and the remote operator workplace of the remote operation control center enable such interaction and transmission of information across the network. Both consist of hardware and software that translate information so that either side can send or receive data and process information to generate actions.

Due to the large volume of data, the vehicle's sensor data must first be encoded by the remote operation modules on the vehicle for transmission over the network and then decoded again by the software of the remote operator workplace on the remote operation control center. Furthermore, information received by the remote operator workplace needs to be displayed in such a way that the remote operator can process it purposefully. This may require modifications by the remote operation workplace's software. Such modifications at the remote operator workplace may also depend on how the vehicle's remote operation modules have modified the respective outgoing data.

Additionally, according to human information processing, the remote operator needs devices to generate input for the vehicle. Such input devices could be, e.g. a mouse, buttons, or a steering wheel. This input data may again be modified by remote operation software at the remote operator workplace and/or the vehicle. Such modification is necessary so that information received from the remote operator workplace is translated into a format that the DAS can process. Thus, the operator's input must be encoded by the remote operator workplace on the remote operation control center side for transmission over the network and then decoded by the remote operation modules on the vehicle side.

The flow of information between design-open remote operation modules within the vehicle and the remote operator workplace described so far does not indicate whether the remote operator's actions constitute remote driving or remote assistance.

### 3.3.2 Differentiation of Remote Assistance and Remote Driving

The decisive criterion for differentiation between remote assistance and remote driving is how information transmitted from the remote operation control center to the vehicle is processed on the vehicle side. In this regard, the type or amount of information is not the crucial element. This focus on vehicle-side processing is necessary because remote operation and human support bridge the current limits of the DAS, particularly where individual DAS modules are not fully functional or unreliable. Accordingly, the remote operation modules on the vehicle side extract the relevant information from the remote operator's input and integrate it into the DAS's information processing workflow.

The DAS receives information from the remote operator and processes the information on either the Sense, Plan or Act stage. When information is processed only on the "Sense" or "Plan" stage, the remote operation constitutes *remote assistance*. In case of remote assistance, the vehicle's DAS still executes the DDT and the remote operator only performs the role of a remote assistant. Only when information is processed on the "Act" stage, for example on controllers or actuators directly, the remote operation constitutes *remote driving*. Accordingly, in that case, the remote operator executes the DDT and acts as a remote driver. In both cases, the safety concept on the vehicle side safeguards the remote operation use case, and the safety concept on the remote operation control center side safeguards remote operators' work and working conditions to allow for safe remote operation.

This also means that the design of the remote operator workplace is not indicative of whether a use case constitutes remote assistance or remote driving. For example, the remote operator workplace could be equipped with control elements similar to those available to an in-vehicle driver, such as steering wheel and pedals. Nevertheless, the use of such control elements does not necessarily imply remote driving. Consider a situation in which the vehicle detects an uninterpretable object ahead and only needs to know whether it is safe to drive over it, for example a shadow or road marking incorrectly classified as an obstacle. In this case, the remote operator's interaction with the steering wheel and pedals may indicate how the situation should be interpreted – for example whether the remote operator would steer around the object or continue driving over it. The remote operation modules on the vehicle side could extract this relevant information from the remote operator's input without the remote operator actually moving or directly controlling the vehicle. This example illustrates that remote assistance can, in principle, be provided through steering wheel and pedal inputs, provided that these inputs are processed as contextual information rather than as direct driving commands on the vehicle side.

Accordingly, we distinguish remote assistance and driving from the perspective of the support or information required by the DAS / vehicle. Based on this, we define the two concepts as follows.

*Remote assistance* describes a form of remote operation where a vehicle receives information from a remote operation control center that is processed on the DAS' Sense and/or Plan stages, in order to facilitate trip continuation, while the vehicle's DAS executes the DDT. The information provided by the remote operation control center is thus not processed on the Act stage of the DAS' information processing (i.e. actuators or controller) architecture and the DAS executes the DDT on its own. Remote Assistance is provided in response to an event only. Events can either be (1) triggered by the DAS which requires information (e.g. unsure object detection, ODD limit reached) or (2) determined by the remote assistant who recognizes that additional information is required (e.g. proactive deactivation by remote assistant in case of a vehicle defect).

*Remote driving* describes a form of remote operation where the vehicle receives information from a remote operation control center that is processed on the Act stage of the DAS' information processing (i.e. actuators or controller) architecture. In this case, the remote driver executes the DDT (Figure 1)[1].

The definitions highlight the major distinction criterion: At which stage of technical information processing of the vehicle is the information received from the remote operation control center processed? Only if the information is directly processed at the Act stage (actuators, controllers) of the DAS, remote driving is provided. When the information is processed at the Sense or Plan stage of the DAS, the use case constitutes remote assistance.

[1] Additional sustained Advanced Driver Assistance Systems (ADAS) can still be active, e.g. Adaptive Cruise Control (ACC) and lane-keeping assist.

### 3.3.3 Characterization of Remote Assistance and Remote Driving Along Frequently Discussed Dimensions

In the following, we describe how the differences between remote driving and remote assistance, as defined in our framework, manifest in practice along the frequently used dimensions in terminology and standards in research, standardization and legal context (see section 2). The aim is to show how the distinction based on the interaction between information processing at the remote operation control center and on the vehicle can be situated and described within these dimensions. Some of these dimensions are directly derived from the definitions of the reviewed literature, while others, are further specified here as implications of the proposed framework for safety-relevant constellations. The table therefore does not introduce additional definition criteria or generally applicable operational requirements, but characterizes how the proposed distinction may become relevant in practice.

***Table 1:*** *Manifestations of differences between remote driving and remote assistance along frequently discussed dimensions*

| **Dimensions** | **Remote Assistance** | **Remote Driving** |
|---|---|---|
| Trigger, Initiator and Frequency | Event-based:<br>(1) Upon request by the DAS (reactive remote assistance), or<br>(2) If the remote operator identifies the need for support during monitoring (proactive remote assistance) | Event-based:<br>(1) Upon request by the DAS (reactive remote driving), or<br>(2) If the remote operator identifies the need for support during monitoring (proactive remote driving)<br><br>Continuous:<br>(3) Continuous remote driving (from starting point to final destination) |
| Driving-related “Act” layer / Execution of the DDT | DAS | Remote driver |
| Driving-related “Sense” layer | (1) Vehicle’s DAS receives support by remote assistant, or<br>(2) Vehicle’s DAS and remote assistant perceive the environment independently of one another, or<br>(3) Remote assistant receives support by the vehicle‘s DAS, e.g., vehicle’s detected objects | Remote driver may receive support by the vehicle’s systems, e.g. vehicle is still able to detect its surroundings and check whether conditions for activating a DAS are met |
| Driving-related “Plan” layer | Remote assistant and vehicle may contribute to Plan layer, and inputs from the remote operator are still subject to verification, replanning, or override by the DAS (i.e., indirect execution of the DDT) | (1) Remote driver performs the DDT incl. planning and executing the planned trajectory (vehicle systems may check whether conditions for activating a DAS are met),<br>or |
| | | (2) Remote driver and vehicle contribute to Plan layer, while the vehicle proposes a path, which may be visualized within the operator's camera feed, and the remote operator then follows it manually. Thus, the vehicle provides a recommendation, which the operator does not explicitly select but executes, allowing for deviations when necessary. |
| Remote operators' input devices | - Input devices are not specified in this article<br>- Various input devices may be used<br>- Input devices could be analogous to those used in remote driving | - Input devices are not specified in this article<br>- Various input devices may be used<br>- Input devices could be analogous to those used in remote assistance |
| Remote monitoring | <u>(1) Outside DAS' ODD:</u><br>Remote monitoring is mandatory when the remote assistant directs the vehicle into an (unfamiliar) area where the DAS would not have entered based in its own information processing, leading the vehicle to operate outside the DAS' ODD. For example, when the remote assistant informs the DAS about a drivable area, prompting it to proceed despite having previously detected an object in that area. In such cases, the remote assistant's input may lead the DAS to ignore a previously detected object, potentially compromising collision avoidance by the DAS. Continuous monitoring is therefore required to promote safety.<br><u>(2) Within DAS' ODD:</u><br>Remote monitoring is not mandatory when the vehicle operates within the DAS' ODD. | Sustained remote monitoring is part of the remote driving task and, as such, is mandatory for performing remote driving. |
| Collision avoidance | <u>(1) Outside DAS' ODD:</u><br>The DAS cannot avoid collision reliably when the vehicle is directed into an unfamiliar area, i.e., outside the DAS' ODD, where the DAS' capabilities are insufficient to ensure safe operation.<br><u>(2) Within DAS' ODD:</u> | Remote driver avoids collisions and can be supported by integrated safety concepts or remote operation software functionalities. |

|  | The DAS avoids collisions when operating within its ODD. |  |
|---|---|---|

Crucially, the dimensions listed in the table above show where differences between remote assistance and remote driving manifest. They can be traced back to the different interactions between the vehicle and the remote operator, as illustrated in Figure 1, and the derived definitions of remote assistance and remote driving (section 3.3.2). As such, the design of the remote operator's workplace is not indicative for the definition of remote assistance and remote driving. Rather, the task that needs to be performed influences the design of the remote operator's workplace (e.g. Gary et al., 2025).

Furthermore, remote monitoring is not described as a separate, distinctive form of remote operation because effective human monitoring depends on a defined objective, requiring clarity on what the operator must detect, identify, or react to. As such, remote monitoring is part of remote assistance and remote driving. If the remote driver performs remote driving and thus executes the DDT, monitoring the vehicle motion is mandatory for the remote driver. In contrast, in the case of remote assistance, the remote assistant may issue a maneuver release and then voluntarily monitor the vehicle completing the maneuver, as long as the DAS remains within its ODD.

# 4 Discussion

For application in various academic disciplines and for different stakeholders, we propose an understanding of remote operation that is based on information processing of the vehicle and the human remote operator. The two forms of remote operation (remote driving and remote assistance) are compared with definitions in recent research and practice. Since, at the time of writing, Germany has enacted regulations governing the remote operation of on-road vehicles, we then demonstrate the broad applicability of our differentiation by applying it exemplarily to the German legal framework. Finally, an outlook is provided on open questions and future research needs related to the proposed differentiation, its practical implementation and its role within broader socio-technical and regulatory contexts of remote operation.

## 4.1 Situating the Proposed Definitions in Current Research and Practice

It should be noted that different documents take different perspectives on remote operation leading to definitions of key terms with specific focuses. For instance, SAE J3016 (2021) views remote operations against the background of increasing driving automation. In contrast, BSI Flex 1887 (2024) specifically addresses "remote operations in support of tests and trials for the early-stage deployment of automated vehicle technologies and capabilities" (p. iv). The technical report by 5GAA (2021) again aims at business considerations for remote operation (or tele-operation). Then there are legal documents such as EU 2022/1426 and the German StVG, AFGBV and StVFernLV providing legal frameworks for remote operation to take place in public road traffic. These different perspectives should be kept in mind when viewing the respective definitions. As such, this article does not seek to invalidate previously proposed definitions and conceptualizations.

Rather, our framework contributes to current research and practice through a more in-depth examination of the teleoperation process by identifying a common basis derived from discussions among experts with both human-centered and technical perspectives from different disciplines, including academia, teleoperation service providers, and other industry stakeholders. We propose a detailed differentiation of remote driving and remote assistance by considering the level of detail that is relevant in order to truly distinguish the terms from one another and by specifying how information is processed in the interaction between a vehicle and a remote operation control center.

For discussions on remote operation, we recommend to clearly specify how the information provided by the remote operation control center is to be utilized during the technical information processing on the vehicle side (see Figure 1). In light of multiple perspectives and by adopting an appropriate level of abstraction, our framework supports a more holistic understanding of different types of remote operation, without depending on a specific technical implementation. At the same time, it captures the underlying process and the relevant forms of interaction between the remote operation control center and the vehicle-side DAS.

In this way, a given remote operation concept can be specified with regard to crucial aspects early on, such as who receives and processes which information, who is performing the DDT at a given point in time, which safety concepts are suitable and what the remote operator workplace includes. Our framework helps clarify the assignment of DDT performance between the vehicle's DAS and the remote operator, including changes in this assignment over time. This can be helpful for discussions concerning downstream issues, such as the arrangement of legal requirements for remote operation.

This article contributes to current research by providing a link between activities of the remote operator and the operated vehicle through the lens of information processing on both the human operator side and the vehicle side. Thereby, we specifically point out that

(1) the actions performed by the remote operator do not determine whether the activity constitutes remote assistance or remote driving,
(2) the design of the remote operator workplace does not determine whether the activity constitutes remote assistance or remote driving,
(3) remote assistance and remote driving are defined by the information processing stage at which the vehicle's DAS processes the incoming information from the remote operation control center.

## 4.2 Exemplary Application of the Definitions and Framework to the Legal Situation in Germany

In Germany, in 2021 the German Road Traffic Act (German original: "Straßenverkehrsgesetz"; abbr. StVG) was amended by including the sections 1d to 1l to cover vehicles operated by a Level 4 DAS in regular operation on public roads. Alongside with this came the new role of a "technical supervisor" (German original: "technische Aufsicht") that corresponds to the remote operator. Within the German Road Traffic Act, the roles of a vehicle operated by Level 4 DAS on the one hand and the technical supervisor on the other hand, are differentiated from each other by assigning specific tasks. The role of the technical supervisor has been further detailed in 2022 in the ordinance regulating authorization and operation of autonomous vehicles (German original: "Autonomes Fahren Genehmigungs- und Betriebsverordnung"; abbr. AFGBV). The role of a technical supervisor, carried out by a natural person, fulfils additional obligations assigned to the vehicle keepers of vehicles operated by a Level 4 DAS. Issues of maintenance and safe operation of the Level 4 DAS are covered by this role.

Within the German Road Traffic Act, the roles of a vehicle operated by Level 4 DAS on the one hand and the technical supervisor on the other, are differentiated from each other by assigning specific tasks. The vehicle must be able to execute the DDT on its own. The vehicle capabilities shall cover the transition to a minimal risk condition whenever the continuation of the journey would require to breach a regulation in the Road Traffic Code. The same applies e.g. when reaching system limits, the boundaries of a licensed area of operation or whenever a technical failure occurs. In this case, the vehicle must - clearly and perceptibly - inform the technical supervisor and provide the necessary information. The technical supervisor then either evaluates the situation and decides upon the release of a maneuver provided by the vehicle or may suggest an alternative driving maneuver. However, it always remains the task of the vehicle to verify maneuvers and potentially refuse execution or perform a minimal risk maneuver if a traffic participant or other person might be put at risk (Sec. 1e para. 2 StVG). The technical supervisor is also responsible for evaluating signals on function status provided by the vehicles' technical equipment and taking appropriate measures to secure vehicles operated by a Level 4 DAS that transitioned to a minimal risk state condition. The technical supervisor might also deactivate the vehicle operated by Level 4 DAS and act in the event of system errors. Generally, the technical supervisor is not required to continuously monitor the vehicle, but act whenever the Level 4 DAS requests support.

In addition to the responsibilities already described in the German Road Traffic Act, further details regarding requirements and proceedings for the technical supervisor are specified in Sec. 14 of the ordinance (AFGBV). Here, three main topics are addressed: qualification, delegation of tasks to other staff and the investigation of the trigger for a minimal risk maneuver as well as assessment of necessity for these maneuvers. In cases of minimal risk maneuvers, the technical supervisor shall investigate the trigger and the necessity thereof before leaving this state (including documentation). Further tasks in connection with the recovery of the vehicle are also assigned to the technical supervisor that are, however, carried out on site and not remotely.

Since the vehicle's Level 4 DAS is always required to perform the DDT, the "technical supervisor" as defined in the StVG represents a remote assistant within our framework. The information provided by the "technical supervisor" in response to the vehicle's request must not be used on the Act stage of the vehicle's information processing, but on the Sense and Plan stages, because the vehicle's capabilities to evaluate the maneuver proposal of the remote assistant must be still in place. Furthermore, as outlined in Table 1, acting upon a vehicle's request constitutes event-based reactive remote assistance.

In addition to remote assistance, Germany has also allowed remote driving on public roads since December 2025, with the newly enacted Road Traffic Remote Driving Ordinance (Straßenverkehr-Fernlenk-Verordnung, abbrev. StVFernLV) which is limited to a period of 5 years.

### 4.3 Outlook and future work

Based on human and technical information processing and their interaction in remote operation concepts, we propose a joint understanding and terminology for remote driving and remote assistance. The proposed framework is intended to be applicable and useable across different stakeholders and disciplines and to support interdisciplinary discussions among actors in science, industry, policy and regulatory authorities. Our focus lies on the flow of information between remote operation control center and the vehicle using a wireless network. The key criterion to distinguish between remote driving and remote assistance is how the DAS processes information provided by the remote operator. By focusing on human and technical information processing, the framework and detailed definitions keep a design-neutral stance and provide a basis for practitioners in industry, regulation, and policy makers to clarify next steps towards implementation of remote assistance and remote driving on public roads.

At the same time, several aspects remain outside the scope of this article and require further investigation. First, it needs to be clarified if and how passengers of the operated vehicles should be informed about the different types of remote operations affecting their journey. Second, our focus on driving-related information processing excludes other remotely executed tasks, such as dispatching or fleet-management activities. Third, this article does not address the skills, qualifications, and training requirements for remote operators, nor the extent to which these may differ between remote driving and remote assistance. Such differences may arise from variations in information flow as well as from the distinct tasks performed by remote drivers and remote assistants. Furthermore, remote operation may be understood as part of a broader socio-technical system that also considers general traffic (flow), societal needs and general working conditions of remote operators. For a detailed compilation of open research questions, we refer to the report of the Working Group "Research Needs in Teleoperation" (2025).

## Acknowledgements

This work was funded by the joint research project "STADT:up - Concepts and Pilot Applications for Continuous Automated Driving in Urban Areas" and SAFESTREAM. The projects are supported by the German Federal Ministry for Economic Affairs and Energy (BMWE), based on a decision of the German Bundestag. The authors are solely responsible for the content of this publication.

## References

5GAA. (2021). *Tele-Operated Driving (ToD): Tele-operated Driving Use Cases, System Architecture and Business Considerations: Technical Report*.

Amador, O., Aramrattana, M., & Vinel, A. (2022). A Survey on Remote Operation of Road Vehicles. *IEEE Access*, *10*, 130135–130154. https://doi.org/10.1109/ACCESS.2022.3229168

Aramrattana, M., Bekmecheri, N., Cortés, A., La Fortelle, A. de, Diermeyer, F., Hallgarten, M., Hornauer, S., Jansson, J., Lauer, M., Li, J., Naumann, M., Oehl, M., Quintero, K., Rehder, E., Schrank, A., Stiller, C., Tomizuka, M., Vanzura, M., Visishta, P., . . . Zhan, W. (2025). *Advancing Automated Driving in Highly Interactive Scenarios through Behavior Prediction, Trustworthy AI, and Remote Operations.* 36th IEEE INTELLIGENT VEHICLES SYMPOSIUM (IV 2025). https://ieee-iv.org/2025/wp-content/uploads/sites/3/2024/12/Advancing-Automated-Driving-in-Highly-Interactive-Scenarios-through-Behavior-Prediction.pdf

Bogdoll, D., Orf, S., Töttel, L., & Zöllner, J. M. (2022). Taxonomy and Survey on Remote Human Input Systems for Driving Automation Systems. In K. Arai (Ed.), *Lecture Notes in Networks and Systems. Advances in Information and Communication* (Vol. 439, pp. 94–108). Springer International Publishing. https://doi.org/10.1007/978-3-030-98015-3_6

Brandt, T., Wilbrink, M., & Oehl, M. (2024). Transparent internal human-machine interfaces in highly automated shuttles to support the communication of minimal risk maneuvers to the passengers. *Transportation Research Part F: Traffic Psychology and Behaviour*, *107*, 275–287. https://doi.org/10.1016/j.trf.2024.09.006

Brecht, D., Gehrke, N., Kerbl, T., Krauss, N., Majstorović, D., Pfab, F., Wolf, M.-M., & Diermeyer, F. (2024). Evaluation of Teleoperation Concepts to Solve Automated Vehicle Disengagements. *IEEE Open Journal of Intelligent Transportation Systems*, *5*, 629–641. https://doi.org/10.1109/OJITS.2024.3468021

The British Standards Institution (2024). *Human factors for remote operation of vehicles – Guide* (BSI Flex 1887).

Bubb, H., Vollrath, M., Reinprecht, K., Mayer, E., & Körber, M. (2015). Der Mensch als Fahrer. In H. Bubb, K. Bengler, R. E. Grünen, & M. Vollrath (Eds.), *Automobilergonomie* (pp. 67–162). Springer Fachmedien Wiesbaden. https://doi.org/10.1007/978-3-8348-2297-0_3

Chehwan, P. (2025, April 19). *Beti, PTO to cover white spots of mobility in rural areas with AV.* CCAM Partnership. EUCAD Symposium 2024, Dublin. https://www.connectedautomateddriving.eu/wp-content/uploads/2024/04/EUCAD2024_RemoteManagement_2.beti_Pierre_Chehwan.pdf

Finland, Germany, and the United Kingdom. (2024, January 3). *Remote activities related to driving* (UNECE/TRANS/WP.1/2024/3). https://unece.org/sites/default/files/2024-02/ECE-TRANS-WP1-2024-3e.pdf

Gary, S., Maag, C., Merkel, N., & Neukum, A. (2025). *Psychologisch-technische Anforderungen an einen Teleoperatorarbeitsplatz*. https://doi.org/10.60850/bericht-f167

Karle, P., & Lienkamp, M. (2021). *Autonomous Driving Software Engineering - Lecture 01: Introduction to Autonomous Driving*. https://doi.org/10.13140/RG.2.2.10596.76165

Majstorovic, D., Hoffmann, S., Pfab, F., Schimpe, A., Wolf, M.-M., & Diermeyer, F. (2022). Survey on teleoperation concepts for automated vehicles. In *2022 IEEE International Conference on Systems, Man, and Cybernetics (SMC)* (pp. 1290–1296). IEEE. https://doi.org/10.1109/SMC53654.2022.9945267

Mozaffari, S., Al-Jarrah, O. Y., Dianati, M., Jennings, P., & Mouzakitis, A. (2022). Deep Learning-Based Vehicle Behavior Prediction for Autonomous Driving Applications: A Review. *IEEE Transactions on Intelligent Transportation Systems*, *23*(1), 33–47. https://doi.org/10.1109/tits.2020.3012034

On-Road Automated Driving (ORAD) committee (2021). *Taxonomy and Definitions for Terms Related to Driving Automation Systems for On-Road Motor Vehicles* (SAE J3016). SAE International. https://www.doi.org/10.4271/J3016_202104

Parasuraman, R., Sheridan, T. B., & Wickens, C. D. (2000). A model for types and levels of human interaction with automation. *IEEE Transactions on Systems, Man, and Cybernetics - Part a: Systems and Humans, 30*(3), 286–297. https://doi.org/10.1109/3468.844354

Prakash, J., Vignati, M., & Sabbioni, E. (2023). Performance of Successive Reference Pose Tracking vs Smith Predictor Approach for Direct Vehicle Teleoperation under Variable Network Delays. *IEEE Transactions on Vehicular Technology*, 1–10. https://doi.org/10.1109/TVT.2023.3339877

Prakash, J., Vignati, M., Vignarca, D., Sabbioni, E., & Cheli, F. (2023). Predictive Display With Perspective Projection of Surroundings in Vehicle Teleoperation to Account Time-Delays. *IEEE Transactions on Intelligent Transportation Systems*, *24*(9), 9084–9097. https://doi.org/10.1109/TITS.2023.3268756

Schrank, A., Walocha, F., Brandenburg, S., & Oehl, M. (2024). Human-centered design and evaluation of a workplace for the remote assistance of highly automated vehicles. *Cognition, Technology & Work.* Advance online publication. https://doi.org/10.1007/s10111-024-00753-x

Schrank, A., Wilbrink, M., Brandenburg, S., & Oehl, M. (2025). Improving road user perception in adverse weather: An augmented human–machine interface for remote assistants of automated vehicles. *Transportation Research Part F: Traffic Psychology and Behaviour*, *113*, 500–516. https://doi.org/10.1016/j.trf.2025.05.017

Schwindt, S., Heller, A., Theobald, N., & Abendroth, B. (2023). Who Will Drive Automated Vehicles? - Usability Context Analysis and Design Guidelines for Future Control Centers for Automated Vehicle Traffic. In *AHFE International, Human Factors in Transportation.* AHFE International. https://doi.org/10.54941/ahfe1003796

Schwindt-Drews, S., & Abendroth, B. (2025). Design Recommendations for Operation Center HMIs in Automated Fleet Operations: A Field Study from the Campus FreeCity Project. In *Uni-DAS e.V. Workshop Fahrerassistenz und automatisiertes Fahren 2025.* https://www.uni-das.de/images/pdf/fas-workshop/2025/FAS2025-07-Schwindt-Drews.pdf

Schwindt-Drews, S., Frenzel, J.-P., Hans, O., Abendroth, B., & Adamy, J. (2025). Competencies and Technological Support for Remote Operator of Automated Vehicles: A Literature Review and Perspectives. In S. Jin, J. H. Kim, Y.-K. Kong, J. Park, & M. H. Yun (Eds.), *Springer Series in Design and Innovation. Proceedings of the 22nd Congress of the International Ergonomics Association, Volume 4. IEA 2024.* (Vol. 56, pp. 187–192). Springer Nature Singapore. https://doi.org/10.1007/978-981-95-0289-9_27

Straßenverkehrsgesetz in der Fassung der Bekanntmachung vom 5. März 2003 (BGBl. I S. 310, 919), das zuletzt durch Artikel 2 des Gesetzes vom 12. Mai 2026 (BGBl. 2026 I Nr. 142) geändert worden ist. https://www.gesetze-im-internet.de/stvg/

Tener, F., & Lanir, J. (2023). Investigating intervention road scenarios for teleoperation of autonomous vehicles. *Multimedia Tools and Applications.* Advance online publication. https://doi.org/10.1007/s11042-023-17851-z

Teng, S., Hu, X., Deng, P., Li, B., Li, Y., Ai, Y., Yang, D., Li, L., Xuanyuan, Z., Zhu, F., & Chen, L. (2023). Motion Planning for Autonomous Driving: The State of the Art and Future Perspectives. *IEEE Transactions on Intelligent Vehicles, 8*(6), 3692–3711. https://doi.org/10.1109/tiv.2023.3274536

Verordnung über Ausnahmen von straßenverkehrsrechtlichen Vorschriften für ferngelenkte Kraftfahrzeuge (Straßenverkehr-Fernlenk-Verordnung - StVFernLV). https://www.gesetze-im-internet.de/stvfernlv/

Verordnung zur Genehmigung und zum Betrieb von Kraftfahrzeugen mit autonomer Fahrfunktion in festgelegten Betriebsbereichen (Autonome-Fahrzeuge-Genehmigungs-und-Betriebs-Verordnung - AFGBV). https://www.gesetze-im-internet.de/afgbv/

Walocha, F., Schrank, A., Nguyen, H. P., & Ihme, K. (2025). Multimodal Assessment of Mental Workload During Automated Vehicle Remote Assistance: Modeling of Eye-Tracking-Related, Skin Conductance, and Cardiovascular Indicators. *Information*, *16*(1), 64. https://doi.org/10.3390/info16010064

Wolf, M.-M., Krauss, N., Schmidt, A., & Diermeyer, F. (2025). Control Center Framework for Teleoperation Support of Automated Vehicles on Public Roads. In *2025 IEEE Intelligent Vehicles Symposium (IV)* (pp. 2538–2545). IEEE. https://doi.org/10.1109/IV64158.2025.11097634

Working Group “Research Needs in Teleoperation”. (2025). *Technical report of the working group “Research Needs in Teleoperation”*. www.bast.de/teleoperation-en / https://doi.org/10.60850/bericht-f166b

Zhao, L. (2023). *Teleoperation and the influence of driving feedback on drivers’ behaviour and experience.* KTH Royal Institute of Technology, Stockholm. https://kth.diva-portal.org/smash/record.jsf?pid=diva2%3A1754796&dswid=9223

Zhao, L., Nybacka, M., & Rothhämel, M. (2023). A Survey of Teleoperation: Driving Feedback. In *2023 IEEE Intelligent Vehicles Symposium (IV)* (pp. 1–8). IEEE. https://doi.org/10.1109/IV55152.2023.10186553